\documentclass[twocolumn,prl,aps,superscriptaddress,showpacs]{revtex4-1}

\usepackage{mathptmx}
\usepackage{subfigure}
\usepackage{dcolumn}
\usepackage{amsmath,amssymb}
\usepackage{bm}
\usepackage{color}
\usepackage{latexsym}
\usepackage{epstopdf}
\usepackage{color}
\usepackage[english]{babel}
\usepackage{latexsym}

\usepackage{psfrag,graphicx} 
\usepackage{epsf} 
\usepackage{subfigure} 
\usepackage{amsmath} 
\usepackage{amssymb} 
\usepackage{amsfonts}
\usepackage{bm}
\usepackage{natbib}
\usepackage{epstopdf}
\usepackage{appendix}

\definecolor{mygrey}{gray}{0.35}
\definecolor{myblue}{rgb}{0.2,0.2,0.8}
\definecolor{myzard}{cmyk}{0,0,0.05,0}
\definecolor{mywhite}{rgb}{1,1,1}
\definecolor{mywhite}{rgb}{1,1,1}
\definecolor{myred}{rgb}{1,0.,0.3}

\usepackage[colorlinks=true,citecolor=myblue,linkcolor=myred]{hyperref}

\def\be{\begin{equation}}
\def\ee{\end{equation}}
\def\ba{\begin{align}}
\def\enda{\end{align}}
\def\bi{\begin{itemize}}
\def\ei{\end{itemize}}

\def\dd{\mathord{\rm d}} 
 \def\ee{\mathord{\rm e}}
 
 \def\ii{\mathord{\rm i}}

\def\dd{\mathord{\rm d}} 
 \def\ee{\mathord{\rm e}}
 
 \def\ii{\mathord{\rm i}}

\renewcommand{\ii}{{\rm i}}
\renewcommand{\ee}{{\rm e}}

\def\beq{\begin{equation}}
\def\beq{\begin{equation}}
\def\eeq{\end{equation}}

 \newcommand{\ket}[1]{|#1\rangle}

\def\dd{\delta}
\def\one{{\bf 1}}
\def\kk{{\bf k}}

\def\rr{{\bf r}}

\def\LL{{\rm L}}
\def\cc{{\rm c}}
\def\dd{{\rm d}}

\def\ii{{\bf i}}
\def\jj{{\bf j}}

\begin{document}


\title{Synthetic Gauge Fields for the Vibrational Excitations of Trapped ions}

\author{Alejandro Bermudez}
\affiliation{
Institut f\"ur Theoretische Physik, Albert-Einstein Allee 11, Universit\"at Ulm, 89069 Ulm, Germany 
}
\affiliation{
Departamento de F\'isica Te\'orica I,
Universidad Complutense, 
28040 Madrid, 
Spain
}

\author{Tobias Schaetz}
\affiliation{
Max-Planck-Institut f\"ur Quantenoptik, Hans-Kopfermann-Strasse 1, D-85748 Garching, Germany
}
\affiliation{
Albert-Ludwigs-Universit\"at Freiburg, Physikalisches Institut, Hermann-Herder-Str3, 79104 Freiburg, Germany}

\author{Diego Porras}
\affiliation{
Departamento de F\'isica Te\'orica I,
Universidad Complutense, 
28040 Madrid, 
Spain
}


\begin{abstract}
The vibrations of a collection of ions in a microtrap array can be described in terms of tunneling phonons.
We show  that the vibrational couplings may be tailored  by using a gradient of the trap frequencies, together with a periodic driving of the trapping potentials. These ingredients allow us to induce effective gauge fields on the vibrational excitations, such that phonons  mimic the behavior  of  charged particles in a magnetic field. In particular, microtrap arrays are well suited to realize a quantum simulator of the famous Aharonov-Bohm effect, and observe the paradigmatic edge states typical from quantum-Hall samples and topological insulators.
\end{abstract}
\maketitle

{\it Introduction.--}
The ultimate goal of  quantum simulation (QS) is to find experimental platforms where quantum many-body physics 
can be explored~\cite{feynman_qs}. 
This challenge  requires efficient methods to prepare quantum states, an exquisite control of the interactions, and precise measurement techniques. Quantum-information technologies have an important application in this context, since they provide us with a powerful toolbox for the manipulation of quantum systems.  In particular, trapped ions~\cite{leibfried03rmp} are an interesting candidate, their main advantage  being  an unrivaled efficiency in preparing and measuring  quantum states at the single-particle level. Besides, the strong long-range Coulomb  interactions make them   suitable for the QS of a variety of collective phenomena, from quantum magnetism~\cite{qs}  to dissipative models~\cite{barreiro}. So far, there have been experiments with up to  $9$ ions~\cite{ising_onset}, and a big effort is being focused on scaling them up. A promising avenue are the so-called  two-dimensional arrays of microtraps (2DAM)~\cite{surfacetraps}, which may open  new routes towards the many-body regime. Unfortunately, this setup still faces some hurdles, such as the large distances  between ions
 ($d_x \approx 40$ $\mu$m) leading to weak spin-spin interactions. In order to realize QS schemes  based on vibration-mediated interactions, it is  fundamental to overcome these issues. Alternatively, focusing directly on the vibrational modes \cite{porras04prl.b}  yields a significant speed-up  with respect to  decoherence rates. In fact, the transfer of vibrational excitations between two aligned traps was recently observed~\cite{exp.microtraps}. 

In this Letter, we show how to tailor the vibrational couplings in a 2DAM, such that this speed-up is exploited. This opens the possibility of building a QS of  lattice bosons under synthetic gauge fields. We note that laser-based methods for neutral atoms might also lead to effective gauge fields~\cite{laser_assisted_ol}.
Our proposal, however, relies  on the different concept of photon-assisted tunneling~\cite{photon_assisted,photon_assisted_ol}, and
requires  a gradient of the individual trapping frequencies, together with a periodic driving of the trapping potentials that can be achieved by an optical  force. 
This Letter is structured as follows:
{\it i)} We show that the amplitude and phase of the vibrational couplings between ions can be tuned by inducing resonances that correspond to the absorption/emission of photons from a classical driving field (photon-assisted tunneling). 
{\it ii)} We extend this result to 2D, and show how it leads to the implementation of synthetic gauge fields, where  phonons move like charged particles in a lattice. 
{\it iii)} We present an implementation of the required drivings by means of optical forces, such that the optical phase can be intrepreted as an effective gauge field.
{\it iv)} We propose a proof-of-principle of our ideas with  four ions in a plaquette displaying a discrete version of the celebrated Aharanov-Bohm effect~\cite{aharonov_bohm}.
{\it v)} We suggest to concatenate those plaquettes in ladders, leading to  Aharonov-Bohm cages~\cite{ab_cages}, and allowing to observe the edge states characteristic of quantum-Hall samples and topological insulators~\cite{ti}.

\begin{figure}

\centering
\includegraphics[width=0.8\columnwidth]{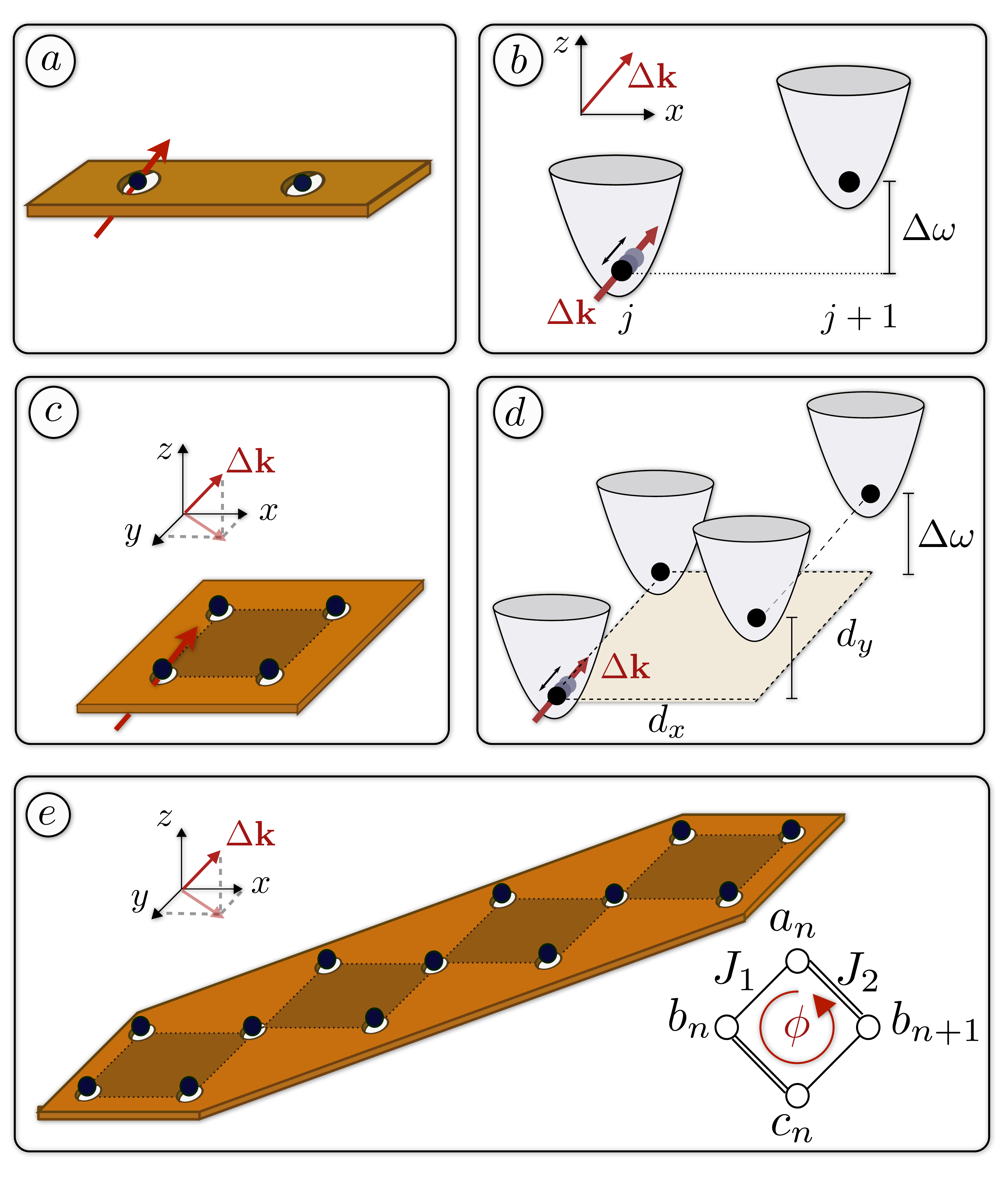}
\caption{ {\bf Arrangement of ion microtraps:}  Schematic representation of the microtrap layout for: a)  two-ion link, c) four-ion plaquette; e) many-ion rhombic ladder  (the relevant parameters of the effective Hamiltonian~\eqref{ab_rhombic} are also shown). Requirements for the photon-assisted tunneling of phonons:  b) The frequencies of two adjacent traps $\omega_1,\omega_2$ are shifted by $\Delta\omega$. By shining a pair of Raman lasers, one assists the phonon transfer; d)  For a plaquette,  the gradient is along  $x$ and the  $\Delta{\bf k}$ has a component along  the $x$-$y$-$z$ axes.}
\label{fig_1}
\end{figure}

{\it i) Photon-assisted tunneling.--} 
We introduce our scheme for two ions with mass $M$ and charge $e$, trapped by independent potentials with frequencies $\omega_{1,2}$ (Figs.~\ref{fig_1}(a)-(b)). The equilibrium positions, separated by $d_x$, lie along the $x$ direction, and the axial vibrational modes are periodically driven.
The  Hamiltonian is $H(\tau) = H_0(\tau) + H_{\rm c}$, with ($\hbar = 1$)
\begin{equation}
H_0(\tau) = \hspace{-0.1cm} \sum_{j=1,2} \hspace{-0.1cm} \omega_j a^\dagger_j a_j + H_{\rm d}(\tau), \hspace{0.2cm}
H_{\rm c} = \frac{e^2}{d_x^3} (\delta x_1 - \delta x_2)^2.
\end{equation}
$H_{\rm c}$ is the Coulomb coupling to second order in the ion displacements, $\delta x_j = (a_j + a^\dagger_j)/\sqrt{2 M \omega_j}$, with $a_j^{\dagger}(a_j)$ phonon creation (annihilation) operators. The periodic driving is
\begin{equation}
H_{\rm d}(\tau) = \sum_{j=1,2} \eta_{\rm d} \ \omega_{\rm d} \cos(\omega_{\rm d} \tau + \phi_j) a^\dagger_j a_j,
\label{H.d}
\end{equation}
where $\omega_{\rm d}$ ($\eta_{\rm d}\omega_{\rm d}$) is the driving frequency (strength), and $\phi_j$  a site-dependent phase. We assume that $\omega_1 = \omega$, $\omega_2 = \omega + \Delta \omega$, and  
$\{ \Delta \omega, \eta_{\rm d} \omega_{\rm d} \} \ll \omega$, namely, both the  frequency difference and driving strength are small perturbations to the trapping frequency. In  absence of driving, the vibrational coupling is
\begin{equation}
H_{\rm c} =  J_{\rm c}  (a^\dagger_2 a_1^{\phantom{\dagger}} + a^\dagger_1 a_2^{\phantom \dagger})  ,
\label{H.c}
\end{equation}
where  $J_\cc = - \beta \omega$, and $\beta = e^2/ M \omega^2 d_x^3$. 
Equation~\eqref{H.c} holds for $|J_\cc| \ll \omega$, such that the cross-terms, $a_1 a_2$, $a_1^\dagger a_2^\dagger$, can be neglected in a rotating-wave approximation (r.w.a.). 
This condition, which is  met for the experiments in
~\cite{exp.microtraps}, 
allows us to interpret  the dynamics as  the tunneling of phonons~\cite{porras04prl.b}.

\begin{figure*}
\centering
\includegraphics[width=1.6\columnwidth]{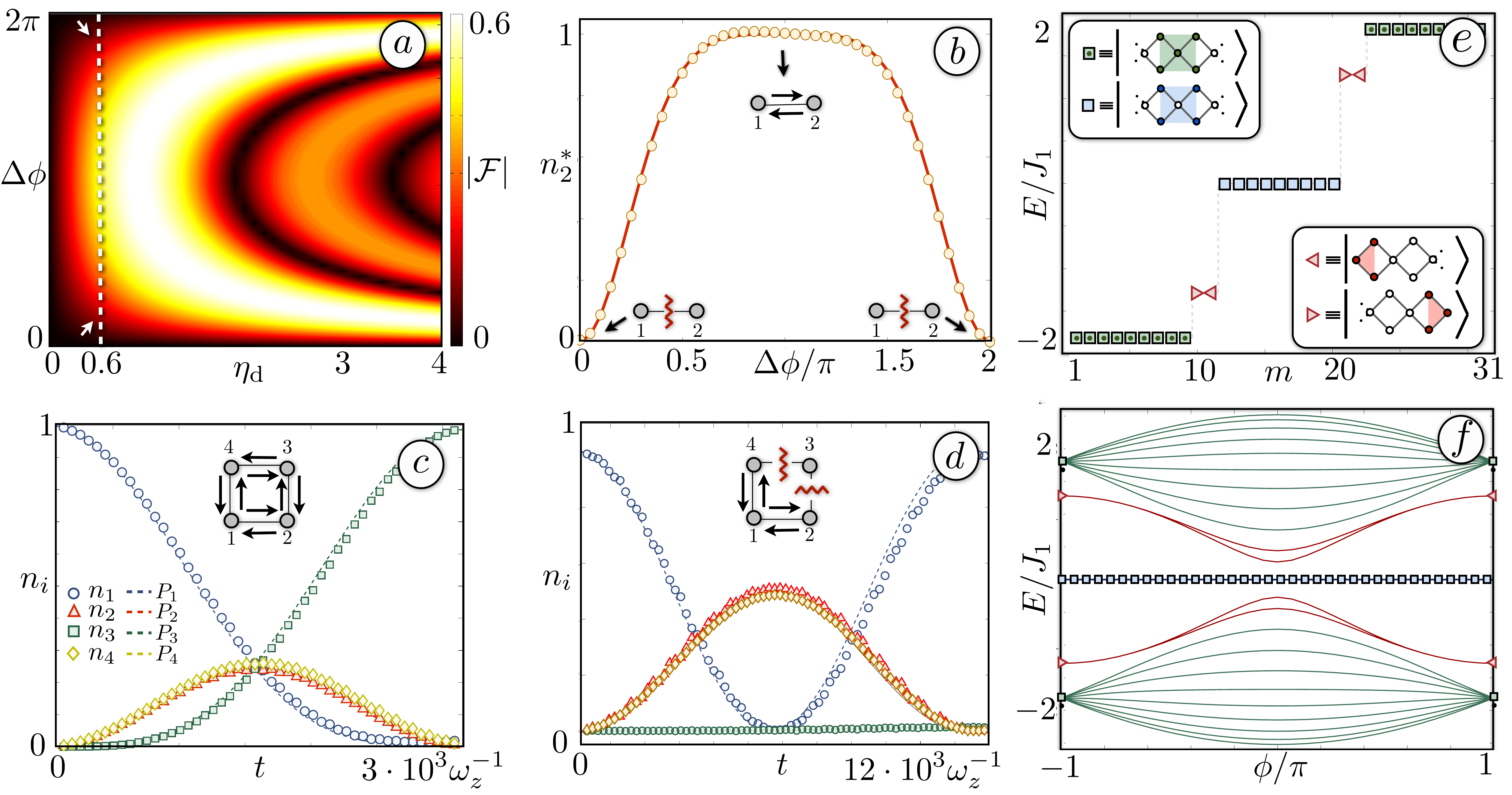}
\caption{ {\bf Photon-assisted tunneling and Aharonov-Bohm cages:}  
(a) Effective hopping amplitude, $|\mathcal{F}_{r=1}|$, as a function of $(\eta_{\rm d},\Delta\phi)$. 
The values of the dashed line at $\eta_{\rm d}\approx 0.6$ are used in (b). 
(b) Photon-assisted hopping for a two-ion link of a single vibrational excitation  $\ket{\psi_0}=a_1^{\dagger}\ket{0}$ under the effective description~\eqref{eff_hopping} (red line), 
and the complete Hamiltonian~\eqref{H.c} with the driving term 
~\eqref{H.L} (yellow dots). 
We plot the vibrational population $n_2^*$ transferred to site 2 after time $t^*=\pi/|J_{\rm [r]}\hspace{-0.5ex}\left( 0.6,\phi \right)|$. 
Maximum phonon transfer occurs at $\Delta\phi=\pi$.
(c,d) Evolution of the phonon excitation in a rectangular plaquette. We plot the phonon populations under approximation~\eqref{eff.tun} ($P_i(t)$), and under the exact Hamiltonian  Eqs.~\eqref{microtrap_driven}-\eqref{H.c.2D} driven by \eqref{H.L} ($n_i(t)$).
In (c) $\phi_{\circlearrowleft}=0$, $\phi_x=\pi,\phi_y=0$, $d_x=d_y|\mathcal{F}_1(\eta_{\rm d},\pi)|^{2/3}$, $n_{\text{max}}=2$, and other parameters same as (b). 
In (d),  $\phi_{\circlearrowleft}=\pi$, and there is an Aharonov-Bohm destructive interference that inhibits tunneling to site 3. $\phi_x=\pi,\phi_y=\pi$, $\Omega_{\LL}=0.25\omega_z$, and other values same as (c).  (e) Energy spectrum $E$ for the $\pi$-flux regime of Hamiltonian \eqref{ab_rhombic} for $N=31$ microtraps. Flat bands appear at  $E\approx\pm2J_1$ (Aharonov-Bohm cages) (see also the schematic description of the eigenstates), and also at $E=0$ (which arise solely due to the geometry of the lattice). Also, in the middle of the gaps, single edge states localized to the boundaries of the ladder arise. (f) Energy level spectrum as a function of the effective flux, which displays a gap-vanishing point at $\phi=0$.}
\label{fig_2}
\end{figure*}

To understand the effects of  driving, we express Eq.\eqref{H.c} in the interaction picture with respect to $H_0(\tau)$, where  $a_i(\tau) = a_i \ee^{-i \omega_i \tau} \ee^{-i \eta_{\rm d} \sin(\omega_\dd \tau + \phi_i)}
\ee^{i \eta{\rm d} \sin(\phi_i)}.$
After the trivial  transformation $a_i \ee^{ i \eta_{\rm d} \sin(\phi_i)} \to a_i$, 
one writes $H_\cc (\tau)$ by replacing the bare Coulomb coupling $J_\cc$  by a time-dependent dressed coupling
\begin{equation}
J(\tau) =  J_\cc \ee^{i \Delta \omega \tau} 
\hspace{-0.3cm} \sum_{s ,s'= -\infty}^{\infty} \hspace{-0.4cm} {\cal J}_s(\eta_{\rm d}) \hspace{-0.1cm}  {\cal J}_{s'}(\eta_{\rm d}) 
\ee^{i s (\omega_{\rm d} \tau + \phi_2)} \ee^{- i s' (\omega_{\rm d} \tau + \phi_1)} ,
\end{equation}
where ${\cal J}_s(\eta_{\rm d})$ are  Bessel functions of the first kind. 
By choosing the driving frequencies  $r \omega_{\rm d} = \Delta \omega$, with $r=1,2,...$, one selects resonant processes that correspond to the absorption/emission of $r$ photons from the classical driving field. 
If  $\Delta \omega \gg J_\cc$, only the resonant terms must be considered after a r.w.a., and the effective vibrational coupling can be written as
\begin{equation}
\label{eff_hopping}
\begin{split}
J_{\rm [r]} \hspace{-0.5ex}\left( \eta_{\rm d},\{\phi_n\} \right)
&= J_\cc \ {\cal F}_r(\eta_{\rm d},\Delta\phi) e^{- i  \frac{r}{2} (\phi_1 + \phi_2)}, \\
{\cal F}_r(\eta_{\rm d},\Delta \phi) &=\hspace{-1ex}\sum
_{s = -\infty}^{\infty}\hspace{-1ex}{\cal J}_s(\eta_{\rm d}) {\cal J}_{s+r}(\eta_{\rm d}) e^{ i (s + \frac{r}{2}) \Delta \phi},
\end{split}
\end{equation}
where $\Delta\phi=\phi_2 - \phi_1$.
Since no perturbative assumption is required on the driving strength, the dressed coupling $J_{\rm [r]}$ may be  close to the bare one $J_\cc$.
Fig. \ref{fig_2}a) shows a calculation of the dressed coupling under different conditions. 
By tuning $\eta_{\rm d}$ and $\{\phi_j\}$, one controls the amplitude and phase of the  tunneling, which may be enhanced or even completely suppressed. 
We note that the coherent control of tunneling is interesting on its own  
\cite{photon_assisted}, and can now be investigated with trapped ions.

{\it ii) Synthetic magnetic fields.--}
We  extend   this scheme to different geometries given by  the ion equilibrium positions in a 2DAM,  separated by $d_x,d_y$, and labeled by vectors of integers, $\ii = (i_x, i_y)$. In the most general situation, trapping frequencies, $\omega_{\alpha, \ii}$, depend on the site $\ii$ and the spatial direction $\alpha = x,y,z$. 
The trap potentials, together with a driving term, are
\begin{equation}
\label{microtrap_driven}
H_{\rm 0}(\tau) = 
\sum_{\ii,\alpha} \omega_{\alpha,\ii} a^\dagger_{\alpha, \ii} a_{\alpha, \ii} + H_\dd(\tau).
\end{equation} 
The vibrational couplings between ions in the array arise due to the Coulomb interaction
\begin{equation}
V_{\rm C} = 
\frac{e^2}{2}\sum_{\ii\neq \jj} \frac{1}{|\rr^0_\ii - \rr^0_\jj + \delta \rr_\ii - \delta \rr_\jj |}, 
\end{equation}
where $\rr^0_\ii$, and
$(\delta \rr_\ii)_\alpha = (a_{\alpha,\ii} + a^\dagger_{\alpha,\ii})/\sqrt{2 M \omega_{\alpha, \ii}}$, 
are the equilibrium positions and relative ion displacements. We assume that  the vibrational modes in different directions are not coupled, and the phonon number is conserved. The validity of this approximation is quantified below. In the harmonic approximation, that is, up to second order in $\delta \rr_\ii$, we find
\begin{equation}
\begin{split}
H_\cc &= \sum_{\ii>\jj,\alpha} 
J_{\cc; \ii,\jj}^\alpha( a^\dagger_{\alpha,\ii} a^{\phantom{\dagger}}_{\alpha, \jj} + a^\dagger_{\alpha,\jj} a^{\phantom{\dagger}}_{\alpha, \ii}),\\
J^{\alpha}_{\cc; {\bf i}, {\bf j}} 
&= - \frac{e^2}{2 M \sqrt{\omega_{\alpha,\ii} \omega_{\alpha,\jj}}}
\frac{3 \big(\rr^0_{\ii-\jj}\big)_\alpha \big(\rr^0_{\ii-\jj}\big)_\alpha -  
|\rr^0_{\ii - \jj}|^2}
{|\rr^0_{\ii - \jj}|^5},
\end{split}
\label{H.c.2D}
\end{equation}
where $\rr^0_{\ii-\jj}=\rr^0_{\ii}-\rr^0_{\jj}$. The assumption of independent vibrations in each direction holds  for 
$|\omega_{\alpha,\ii} - \omega_{\beta,\jj}|_{\alpha \neq \beta} \gg |J^\alpha_{\cc; \ii, \jj}|$, whereas phonon number conservation is valid if 
$\omega_{\alpha,\ii} \gg |J^\alpha_{\cc; \ii, \jj}|$.

In quantum mechanics, charged particles under electromagnetic fields acquire a phase that depends on the field background. To make a QS of this phenomenon, we  focus on the ion motion in  direction $\bar{\alpha}$, and choose a linear gradient along $x$,  $\omega_{\bar{\alpha}, \ii} = \omega_{\bar{\alpha}} + \Delta \omega i_x$, together with phases that depend linearly on the position, $\phi_{{\bf i}} =  \phi_x i_x + \phi_y i_y$. 
Equation (\ref{H.d}) is generalized  to 
$H_\dd(\tau) = \sum_{\ii} \eta_\dd \omega_\dd \cos(\omega_\dd \tau + \phi_\ii ) a_{\bar{\alpha},\ii}^\dagger a_{\bar{\alpha},\ii}$. 
In analogy to the two-ion case, we find that the 
effective vibrational couplings to leading order in $\eta_{\rm d}$~\cite{comment} are the following:
\begin{equation}
J^{\bar{\alpha}}_{{\rm [r]}; \ii, \jj} =
J^{\bar{\alpha}}_{\cc; \ii, \jj} 
{\cal F}_r(\eta_{\rm d},\Delta \phi_{\ii,\jj}) e^{- i \frac{r}{2}(\phi_\ii + \phi_\jj)} \delta_{i_x, j_x + 1}
+
J^{\bar{\alpha}}_{\cc; \ii, \jj} \delta_{i_x, j_x},
\label{eff.tun}
\end{equation}
where $\Delta \phi_{\ii,\jj}=\phi_{\ii}-\phi_{\jj}$, and $ \delta_{i_x, j_x}$ is the Kronecker delta. The first term  describes the photon-assisted tunneling  along $x$, whereas the second one is  the bare coupling  along $y$.  A crucial result is that the amplitude of tunneling around a plaquette, 
\begin{equation}
\nonumber
W_{\circlearrowleft}^{\bar{\alpha}}=J^{\bar{\alpha}}_{{\rm [r]}; \ii,\ii+\boldsymbol{\hat{y}}}   J^{\bar{\alpha}}_{{\rm [r]}; \ii+\boldsymbol{\hat{y}}, \ii+\boldsymbol{\hat{x}}+\boldsymbol{\hat{y}}}     J^{\bar{\alpha}}_{{\rm [r]}; \ii+\boldsymbol{\hat{x}}+\boldsymbol{\hat{y}},\ii+\boldsymbol{\hat{x}}}    J^{\bar{\alpha}}_{{\rm [r]};\ii+\boldsymbol{\hat{x}}, \ii}=|W_{\circlearrowleft}^{\bar{\alpha}}|\ee^{i\phi_{\circlearrowleft}},
\end{equation}
yields an accumulated phase that depends on the laser parameters $\phi_{\circlearrowleft}=-r\phi_y$, and can be recast in terms of the celebrated Aharonov-Bohm phase~\cite{aharonov_bohm},  $\phi_{\circlearrowleft}=e\oint_{\circlearrowleft} {\rm d}\boldsymbol{x}\cdot{\bf A}$, 
where $\oint_{\circlearrowleft}$ is the line integral along the plaquette, and
 ${\bf A} = r\phi_y y/(ed_xd_y)\boldsymbol{\hat{x}}$ is a synthetic vector potential. Accordingly,  phonons move as charged particles  subjected to a magnetic field perpendicular to the microtrap array, and yield a  bosonic counterpart of the  Azbel-Harper-Hofstadter model~\cite{azbel_harper_hofstadter}. We stress that arbitrary fluxes $\phi_{\circlearrowleft}\in[0,2\pi)$ can be attained, even reaching one flux quantum per unit cell;  a regime unaccessible in solid-state materials for realistic magnetic fields. This opens the possibility to observe a dipolar version of the fractal Hofstadter butterfly, among other interesting effects presented in {\it iv)-v)}.

{\it iii)  Realization of the periodic driving.--}
The simplest setup to realize Eq.~\eqref{H.d} would consist of an array of  microtraps, 
where the driving fields are provided by the local  control of the electrodes. 
Since this scheme is yet to be realized and scaled~\cite{surfacetraps,tobiastraps}, we base our alternative approach on state-of-the-art optical forces. 
We focus on  the vibrational modes transverse to the microtrap plane, $\bar{\alpha} = z$,
although  other schemes along the $x$-$y$ plane are equally valid~\cite{note2}. 
We consider  lasers that drive two-photon stimulated Raman
transitions between the electronic levels of the ions
$| 0 \rangle_\ii, | 1 \rangle_\ii$. The lasers are detuned by 
$\omega_{\rm L}$, 
and provide a Raman wavevector $\Delta {\bf k}$,
\begin{equation}
H_{\rm L} = \frac{\Omega_{\rm L}}{2} \sum_\ii O_\ii  \left(
 \ee^{i \Delta {\bf k} (\rr^0_\ii + \delta \rr_\ii) - i \omega_{\rm L} \tau} + {\rm H.c.} \right),
\label{H.L}
\end{equation}
where  $\Omega_{\rm L}$ is the Rabi frequency, and $O_\ii$ is an operator acting on the electronic levels. By a proper choice of the laser detunings and polarizations, one may realize $O_\ii = \mathbb{\one}_\ii$, or other operators like $O_\ii = \sigma^{z}_{\ii}$ that widen the applicability of our QS (Outlook). The effect of the ion-laser interaction can be understood as a periodic driving of the microtrap frequencies under the assumptions below. 
We  consider a gradient along $x$, 
$\omega_{z,\ii} = \omega_z + \Delta \omega i_x$, such that the following conditions are fulfilled 
$ \{ \Delta \omega, \omega_\LL \} 
\ll \{ \omega_\alpha, |\omega_\alpha - \omega_\beta|_{\alpha \neq \beta} \}$. 
Let us define the Lamb-Dicke parameter along $\alpha$, 
$\eta_\alpha = |\Delta {\bf k}_\alpha| / \sqrt{2 M \omega_\alpha}$. 
In the limit $\eta_\alpha \ll 1$, we perform a Taylor expansion of (\ref{H.L}) up to second order in 
$\eta_\alpha$, $H_\LL = H_{\LL,0} + H_{\LL,1} + H_{\LL,2}$. Note that $H_{\LL,0}$ does not affect the vibrational modes, and $H_{\LL,1}$ can be neglected if $\Omega_\LL \eta_\alpha \ll \omega_\alpha$ in a r.w.a. This leads to
\begin{equation}
H_{\LL,2} \approx \Omega_\LL 
\sum_{\alpha,\beta,\ii} \eta_\alpha \eta_\beta 
O_\ii \cos(\Delta \kk \cdot \rr_\ii^0 - \omega_\LL \tau)
a^\dagger_{\alpha,\ii} a_{\beta,\ii}  .
\label{H.L2}
\end{equation}
Finally, by considering 
$|\omega_\alpha - \omega_\beta|_{\alpha \neq \beta} \gg \Omega_\LL \eta_\alpha^2$
\cite{note2}, 
we neglect the  coupling between different directions, and get the announced periodic driving presented in Eq.~\eqref{H.d} with the following identifications, 
$ \omega_\dd=\omega_\LL$, 
$\eta_\dd \omega_\dd = \Omega_\LL \eta_z^2$, and 
$\phi_\ii = - \Delta \kk \cdot \rr_\ii^0$. 

Current microtrap design~\cite{surfacetraps,tobiastraps} is consistent with the above requirements,
$\{\omega_\alpha, \ |\omega_\alpha - \omega_\beta|_{\alpha \neq \beta} \} \gg 
\{ \omega_\LL, \ \Delta \omega \} \gg J_{\ii,\jj} $. 
Typically, $\omega_\alpha/2\pi \approx$ 1-10 MHz, $J/2\pi \approx$ 5 kHz. To fit the inequality, we can take $\omega_\LL/2\pi \approx$ 50 kHz. With a typical Lamb-Dicke parameter of $\eta_\alpha \approx$ 0.2, the condition $\eta_\alpha \Omega_\LL \ll |\omega_\alpha - \omega_\beta|$ 
is still fulfilled \cite{note2}. 
In Fig.~\ref{fig_2}b), we compare the effective description~\eqref{eff_hopping} to the exact optical  forces~\eqref{H.L} for a two-ion array, with  parameters $\Delta\omega= 0.05 \omega_z$, $\eta_{z}=0.2$, $\Omega_{\LL}=0.75\omega_z$,  
$\beta=0.002$, $r=1$, where  the phonon Hilbert space is truncated to $n_{\text{max}}=4$.  We observe an excellent agreement between both descriptions, yielding assisted tunneling for $\Delta\phi\approx\pi$~\cite{photon_assisted}.  

{\it {\it iv)}-{\it v)} Aharonov-Bohm physics in  lattices.--} 
We apply our ideas to  a square lattice, and set $r = 1$,
$|\Delta \kk_x| d_x = 2 \pi n_x + \phi_x$, $|\Delta \kk_y| d_y = 2 \pi n_y + \phi_y$, with $n_x$, $n_y\in\mathbb{Z}$. The latter are introduced because typical ion distances  are larger than optical wavelengths. From Eq. (\ref{eff.tun}), we get the  tight-binding model,
\begin{equation}
H_{\text{eff}} = 
\sum_\ii J^{z}_{[1];\ii,\ii+\boldsymbol{\hat{x}}} a^\dagger_\ii a^{\phantom{\dagger}}_{\ii + \boldsymbol{\hat{x}}} \ee^{-i \phi_{\circlearrowleft}i_y}
+   \sum_{\ii,m>0} J^{z}_{c;\ii,\ii+m\boldsymbol{\hat{y}}} 
a^\dagger_\ii a^{\phantom{\dagger}}_{\ii+m\boldsymbol{\hat{y}}} +\text{H.c.},
\label{tb}
\end{equation}
where $\phi_{\circlearrowleft}=\phi_y$~\cite{note1}. Note that photon-assisted tunneling along the diagonals has been neglected, since for  $\phi_x=2\pi-\phi_y$ that tunneling amplitude  vanishes $\mathcal{F}(\eta_{\rm d},2\pi)=0$ (Fig.~\ref{fig_2}a)). Besides, the remaining diagonal terms,
$J^z_{{\rm [1];\ii,\ii+\boldsymbol{\hat{x}}}+m\boldsymbol{\hat{y}}}$, are negligible for $ m > 1$ due to the fast dipolar decay.

{\it iv) Discrete Aharonov-Bohm effect.--} The simplest realization of this tight-biding model consists of a single plaquette (Figs.~\ref{fig_1}c)-d)). In Figs.~\ref{fig_2}c)-d), we test the validity of the effective dynamics in \eqref{tb}, by comparing with an exact numerical calculation of the complete driven Hamiltonian~\eqref{H.L}.
These results describe a realization of the discrete Aharonov-Bohm effect with minimal required resources. In Fig.~\ref{fig_2}d), we observe that an initial excitation  can follow two possible paths, either  $1\leftrightarrow 2\leftrightarrow 3$ or $1\leftrightarrow 4\leftrightarrow 3$, enclosing a net flux $\phi_{\circlearrowleft} = \pi$. The paths interfere destructively and forbid the phonon to tunnel to site $3$. Conversely, in Fig.~\ref{fig_2}c),  phonons tunnel around the plaquette for $\phi_{\circlearrowleft}=0$.

{\it v) Aharonov-Bohm cages and flatband physics.--} Let us consider an interesting route beyond the single plaquette, which is the rhombic 3-leg ladder presented in Fig.~\ref{fig_1}d). This system  is described by the Hamiltonian
\begin{equation}
\label{ab_rhombic}
H\hspace{-0.05cm}=\hspace{-0.05cm}\sum_{j}J_1(b^{\dagger}_ja_j+c_j^{\dagger}b_{j+1})+J_2(b^{\dagger}_jc_j+\ee^{i\phi}a_j^{\dagger}b_{j+1})+\text{H.c.},
\end{equation} 
where we have labelled the boson operators for each leg as $a_j,b_j,c_j$ (Fig.~\ref{fig_1}e)). This Hamiltonian follows directly from Eq.~\eqref{tb}, when the plaquettes are arranged along a diagonal, with $J_1=e^2/(2md_y^3), J_2=J_1\mathcal{F}_r(\eta_{\rm d},\Delta\phi)(d_y/d_x)^3,$ and $\phi=\phi_y,\phi_x=2\pi-\phi_y$. This model yields two effects. Due to the Aharonov-Bohm interference for $\phi=\pm \pi$, all the modes of the system are localized and one obtains flat vibrational bands (Fig.~\ref{fig_2}e)). In particular, the non-zero energy modes correspond to the so-called Aharonov-Bohm cages, where phonons are not allowed to tunnel two plaquettes apart~\cite{ab_cages}. 
Besides, one finds the so-called edge states, which are mid-gap modes  exponentially localized around the boundaries. By tuning $\phi$, and  $J_2/J_1$, one can explore a transition between two topologically non-equivalent phases (Fig.~\ref{fig_2}d). 

Finally, let us consider the experimental requirements for the implementation of our ideas. 
The duration of a QS to observe the effects of the synthetic gauge fields is of the order of $1/J_{[r]}$, being $J_{[r]}$ of the order of the bare couplings $J_\cc$, which are in the range $1$-$2$ kHz~\cite{exp.microtraps}.  This can be increased to $5$ KHz following the trap design~\cite{tobiastraps}, and even enhanced by orders of magnitude by further miniaturizing the electrode structure, and storing more than one ion per lattice site~\cite{exp.microtraps}.
The main competing decoherence mechanism is heating of the motional modes 
\cite{leibfried03rmp}.
Heating rates as low as $0.07$ phonons /ms have been reported in cryogenic traps \cite{exp.microtraps}, in principle allowing to implement our ideas. Even when heating rates are comparable to couplings $J_{[r]}$, they may induce a thermal background over which propagation of vibrational excitations may still be observed~\cite{exp.microtraps}.
Note that experimental techniques are available for preparation and measurement of phonon states~\cite{meekhof96prl}. Also,  the vibrational spectrum can be measured without local  addressing in  the ions fluorescence sidebands~\cite{leibfried03rmp}.

{\it Conclusions and Outlook.--} 
We have presented a  proposal to induce synthetic gauge fields for ions in microtrap arrays, which is based on the photon-assisted tunneling of vibrational excitations.
By considering trap designs with anharmonicites, effective phonon-phonon interactions can be included~\cite{porras04prl.b}, which may allow us to study strongly correlated phases. Inducing electronic state-dependent drivings, one gets effective spin-orbit couplings that induce disorder \cite{bermudez10njp}. Also, by adding dissipation, i.e. motional heating, one may  study quantum effects in energy transport in the presence of noise \cite{plenio08njp}. These ingredients make a  versatile QS of many-body physics, which would outperform classical computers for $\approx$10 ions, and $\approx 4$ phonons per ion. That size seems feasible in the near future in view of current experimental progress \cite{surfacetraps}.
Finally, our scheme could be extended to other systems such as photons in  arrays of cavities in circuit QED~\cite{schmidt10prb}.

{\it Acknowledgments.-} This work was partially supported by EU STREPs (HIP, PICC), and by QUITEMAD S2009-ESP-1594, FIS2009-10061, CAM-UCM/910758, and RyC Contract Y200200074.


\begin{references}
\bibitem{note2} An effective Raman wave-vector along $\hat{z}$ may be obtained without illuminating the trap surface, by choosing the trapping axis $\hat{z}$ to be tilted with respect  to the surface.
\end{references}
\end{document}